\documentstyle[12pt]{article}
\begin{document}

\tolerance=5000

\def\pp{{\, \mid \hskip -1.5mm =}}
\def\cL{{\cal L}}
\def\be{\begin{equation}}
\def\ee{\end{equation}}
\def\bea{\begin{eqnarray}}
\def\eea{\end{eqnarray}}
\def\tr{{\rm tr}\, }
\def\nn{\nonumber \\}
\def\e{{\rm e}}
\def\D{{D \hskip -3mm /\,}}

\def\SEH{S_{\rm EH}}
\def\SGH{S_{\rm GH}}
\def\AdS5{{{\rm AdS}_5}}
\def\S4{{{\rm S}_4}}
\def\gfv{{g_{(5)}}}
\def\gfr{{g_{(4)}}}
\def\SC{{S_{\rm C}}}
\def\RH{{R_{\rm H}}}

\  \hfill
\begin{minipage}{3.5cm}
%OCHA-PP-??? \\
%NDA-FP-?? \\
%your preprint No. \\
November 2000 \\
%hep-th/yymmxxx \\
\end{minipage}

\vfill

\begin{center}
{\large\bf AdS/CFT correspondence, conformal anomaly and 
quantum corrected entropy bounds}

\vfill

{\sc Shin'ichi NOJIRI}\footnote{nojiri@cc.nda.ac.jp},
and {\sc Sergei D. ODINTSOV}$^{\spadesuit}$\footnote{
On leave from Tomsk State Pedagogical University, 
634041 Tomsk, RUSSIA. \\
odintsov@ifug5.ugto.mx, odintsov@mail.tomsknet.ru}\\

\vfill

{\sl Department of Applied Physics \\
National Defence Academy,
Hashirimizu Yokosuka 239-8686, JAPAN}

\vfill

{\sl $\spadesuit$
Instituto de Fisica de la Universidad de Guanajuato,
Lomas del Bosque 103, Apdo. Postal E-143, 
37150 Leon,Gto., MEXICO }

\vfill

{\bf ABSTRACT}

\end{center}

The role of conformal anomaly in AdS/CFT and related issues is clarified.
The comparison of holographic and QFT conformal anomalies (with account of 
brane quantum gravity contribution) indicates on the possibility for 
brane quantum gravity to occur within AdS/CFT set-up.
3d quantum induced inflationary (or hyperbolic) brane-world is shown to
be realized in frames of AdS3/CFT2 correspondence where the role of 
2d brane cosmological constant is played by effective tension due to 
two-dimensional conformal anomaly. The dynamical equations to describe 
4d FRW-Universe with account of quantum effects produced by conformal
anomaly are obtained. 
The quantum corrected energy, pressure and entropy are found.
Dynamical evolution of entropy bounds in inflationary Universe is estimated
and its comparison with quantum corrected entropy is done.
It is demonstrated that entropy bounds for quantum corrected entropy 
are getting the approximate ones and occur for some limited periods 
of evolution of inflationary Universe.

\newpage

\section{Introduction}

AdS/CFT correspondence \cite{AdS} in its simplest version shows 
remarkable duality between classical higher dimensional gravity 
and (brane) QFT living in less dimensions. The power of dualities 
is explicitly expressed in this new principle. It demonstrates the 
universality of high energy physics in description of areas 
which seem to be quite distant. As it happens usually,
providing new ways in the resolution of well-known problems 
AdS/CFT puts also new questions. Just to mention some of them:
The description of brane quantum gravity in terms of AdS/CFT set-up?
The consistent embedding of Randall-Sundrum orbifold 
compactification as warped compactification in string theory?
AdS/CFT basis for (quantum) entropy bounds origin?
It is expected that holographic conformal anomaly \cite{ca} and its 
counterpart, quantum field conformal anomaly (for a review, 
see \cite{duff}) should play an essential role in the study 
of above as well as related problems.

In the present paper we discuss the role played by conformal 
anomaly in various aspects of AdS/CFT and in related issues. 
First of all, in the next section the comparison of holographic 
and QFT conformal anomalies is done. The account of quantum 
gravity (brane gravity) contribution in quantum conformal 
anomaly is made. As a result, it follows that holographic 
conformal anomaly may be equal to quantum conformal anomaly not 
only for super Yang-Mills theory but also for (non)supersymmetric 
matter theory with brane gravity. This indicates that brane 
(Einstein or Weyl) gravity has some chances to occur in 
AdS/CFT set-up , at least as next-to-leading order effect.

As the other application of conformal anomaly the scenario of 
refs.\cite{NOZ,HHR0} is extended to d3 brane-worlds.
According to this scenario, originally suggested for d5 
brane-worlds, Randall-Sundrum Universe is realized in 
AdS3/CFT2 correspondence as warped compactification (kind of 
holographic RG flow). The essential role is played by 2d conformal 
anomaly (brane cosmological constant is fixed) which introduces 
the effective brane tension. As a result quantum induced 
inflationary or hyperbolic brane-world Universe occurs. 
The details of such construction are described in third section.

In fourth section, again using conformal anomaly
(or anomaly induced effective action) the dynamical evolution 
equations describing 4d FRW-Universe are defined. The same 
equations are obtained by two different formalisms where 
the account of quantum effects in energy/pressure is done. 
It is mentioned that these equations which admit de Sitter 
solution are the basis for anomaly driven inflation which 
is considered to be quite realistic nowdays.
(4d anomaly driven inflation\cite{S} is closely related 
with anomaly driven inflationary brane-worlds\cite{NOZ,HHR0,NOO} 
as it is noted in fourth section). The quantum corrected 
energy and pressure are defined. It is remarkable that quantum 
energy and pressure may include the contributions from quantum 
gravity itself (or in another way, Einstein gravity 
is modified by anomaly induced effective action which accounts 
quantum effects of matter and gravity via the corresponding 
coefficients of conformal anomaly). Following to recent proposal 
by E. Verlinde\cite{EV}  on the study of dynamical evolution of 
entropy bounds \footnote{For earlier discussion of entropy 
in expanding Universe, see, for example\cite{gibbons}.}
the attempt to introduce quantum entropy is done.
Considering de Sitter space as an example it is suggested that large 
quantum contribution to entropy may destroy well-accepted 
entropy bounds (say of Bekenstein-Hawking type). As a result 
the entropy bounds become approximate ones and occur only for 
some time-limited periods of Universe evolution. They also evolve 
 and with the Universe expansion they may appear 
in different way as we speculate.
Brief summary and some outlook is presented in last section.

\section{AdS/CFT and brane quantum gravity}

Let us start from the discussion of AdS/CFT correspondence (on the level 
of holographic conformal anomaly). We argue that 
AdS/CFT set-up opens the window for brane quantum 
gravity as dual theory. In its own turn, this indicates on
even better understanding of Randall-Sundrum 
compactification \cite{RS} within AdS/CFT correspondence 
\cite{AdS} in string theory. 

Let us start from $d+1$ dimensional gravity on AdS background. 
The action is given by
\be
\label{Ii}
S=\int d^{d+1}x \sqrt{-\hat g}\left\{{1 \over 16\pi G}\hat R 
 - \Lambda \right\}\ .
\ee
The AdS metric may be chosen as follows
\be
\label{Iii}
ds^2 = {l^2 \over 4}\rho^{-2}d\rho d\rho 
+ \sum_{i=1}^d \rho^{-1}g_{\mu\nu} dx^\mu dx^\nu
\ee
where $g_{\mu\nu}$ is metric of boundary manifold, and dimensional 
parameter $l$ is introduced explicitly. 
Note that dimensional parameter is related with bulk 
cosmological constant:
\be
\label{Iiii}
l^2 = -{d(d-1) \over 16\pi G \Lambda}
\ee
as it is dictated by Eq. of motion.
Using several methods (holographic RG or expansion 
of classical AdS action in powers of a cut-off parameter), 
one can get the holographic conformal anomaly (CA). 
This CA should correspond to boundary dual QFT. 

Explicitly, for $d=2$:
\be
\label{Iiv}
T=- {l \over \kappa^2}R\ ,\quad \kappa^2 = 16\pi G \ .
\ee
For $d=4$:
\be
\label{Iv}
T=-{l^3 \over 8\kappa^2}\left(G - F\right)
\ee
where $G=R^2 -4 R_{\mu\nu}R^{\mu\nu} 
+ R_{\mu\nu\rho\sigma}R^{\mu\nu\rho\sigma}$, 
$F={1 \over 3}R^2 -2 R_{\mu\nu}R^{\mu\nu}
+ R_{\mu\nu\rho\sigma}R^{\mu\nu\rho\sigma}$ and 
$R$ are $d$-dimensional curvature invariants for 
$d+1$-dimensional theory (\ref{Ii}). 

The next step is comparison with dual QFT CA. 
Explicit calculation gives
\be
\label{Ivi}
T=bF + b'\left(G-{2 \over 3}\Box R\right)
\ee
where
\bea
\label{Ivii}
b&=&{N +6N_{1/2}+12N_1 - 8N_{\rm HD} + 611 N_2 
+ 796 N_W \over 120(4\pi)^2}\nn 
b'&=&-{N+11N_{1/2}+62N_1 -28 N_{\rm HD} + 1411 N_2 
+ 1566 N_W \over 360(4\pi)^2}\ .
\eea
Here $N$, $N_{1/2}$, $N_1$, $N_{\rm HD}$ 
is the number of scalars, (Dirac) spinors, vectors and higher 
derivative conformal scalars which present in boundary QFT. 
$N_2$ denotes the contribution to CA from spin-2 field 
(Einstein gravity) and $N_W$ - the contribution from 
higher derivative Weyl gravity. (For the calculation 
of quantum gravity corrections to CA see \cite{AMM}).  
Note also that $\Box R$-term in 4d CA is ambigious 
as it may depend on the regularization choice. 
Moreover, it may be changed by finite renormalization of classical 
 gravitational action. That is why this term is of no interest in the 
present context.

It is clear that for AdS/CFT correspondence on 
the level of CA, two expressions (\ref{Iv}) and (\ref{Ivi}) 
(when no $\Box R$-term presents) should coincide. To find the 
corresponding QFT dual one has the condition:
\be
\label{Iviii}
b=-b'\ .
\ee
In pure matter sector there is following natural 
possibility:
\be
\label{Iix}
N=6c\ ,\quad N_{1 \over 2}=2c\ ,\quad N_1=c
\ee
where $c$ is an arbitrary number. The choice of $c=N^2 -1$ 
gives ${\cal N}=4$ $SU(N)$ super Yang-Mills theory multiplet
which is known to be conformally invariant theory  
where 
\be
\label{Ix}
b={N^2 -1 \over 4 (4\pi)^2}\ .
\ee
This is standard AdS/CFT correspondence on the level of CAs. 
The AdS/CFT choice
\be
\label{Ixi}
{l^3 \over 8\kappa^2}={N^2 \over 4(4\pi)^2}
\ee
leads to coincidence of holographic conformal 
anomaly with QFT CA in the leading order of large-$N$.

The remark is in order. The choice (\ref{Iix}) is not 
unique. For example, the condition (\ref{Iviii}) is 
fulfilled for 
\bea
\label{Ixii}
&& N_1=2c_1 \ ,\quad N_{1 \over 2}=2 \ ,\quad N=12 \nn
&& N_1=2c_1 \ ,\quad N_{1 \over 2}=4 \ ,\quad N=12 \nn
&& N_1=2c_1 \ ,\quad N_{1 \over 2}=6 \ ,\quad N=5 \ .
\eea
Increasing the number of vectors leads to 
more variants for scalar and spinors. Of course, such theories 
are only conformally invariant ones as free theories (no interaction). 
It is very interesting that condition (\ref{Iix}) in 
matter sector looks as 
\be
\label{Ixiii}
2N+7N_{1 \over 2}=26 N_1\ .
\ee
This relation appears in the study of stability of Nariai 
BH. As it was shown in \cite{NO8}, if conformal GUT matter 
content satisfies $2N+7N_{1 \over 2}>26 N_1$ the 4d Nariai 
BH is stable (in 4d Einstein gravity with quantum 
corrections due to such GUT). 
No BH anti-evaporation occurs. It is also interesting 
that not only SUSY theories satisfy the bound (\ref{Ixiii}), 
but also non-SUSY ones. That suggests the possibility of 
other duals for d5 AdS Einstein gravity. 

What is more interesting, one can take into account the brane QG 
in relation (\ref{Iviii}). The easy  check shows that 
considering Einstein brane gravity ($N_2=1$, $N_W=0$) 
with scalars, spinors and 
vectors, the relation (\ref{Iviii}) holds for Einstein gravity 
with ($N_1=17$, $N_{1 \over 2}=0$, $N=10$), 
($N_1=17$, $N_{1 \over 2}=2$, $N=3$), 
($N_1=18$, $N_{1 \over 2}=2$, $N=16$), 
($N_1=18$, $N_{1 \over 2}=4$, $N=9$), 
($N_1=18$, $N_{1 \over 2}=6$, $N=2$). 
With increase of vector number, the number of 
choice for scalars and spinors increases as well. 
Similarly, it is easy to check that 4d Weyl gravity with some 
matter content gives QFT anomaly 
reproducing holographic anomaly. This indicates to possible 
AdS/CFT duality 
between bulk d5 AdS gravity and brane quantum (Einstein or Weyl)
gravity with some (non)-supersymmetric amount of matter.
Of course, other checks
should be considered in such proporsal as 
only comparison of anomalies is not enough. However, it gives 
useful indication on dual brane QG for d5 bulk gravity!

Moreover, various scenarios can be suggested. 
 For example, QG with matter may be subdominant. 
It gives the contribution to $b$ of about ${7 \over (4\pi)^2}$.
Then if $N>7$, the dual QFT may be ${\cal N}=4$ $SU(N)$ 
super Yang-Mills theory with (Einstein) QG with matter. 
The Einstein gravity with (non-supersymmetric) matter 
will correspond to next-to-leading order 
of large $N$ expansion. Other variants of such scenario may be 
considered as well.

In $d=2$ cases the comparison is even easier. QFT CA is equal 
here:
\be
\label{Ixiv}
T={cR \over 2\pi}\ ,
\ee
where $c=N+{N_{1 \over 2}} - 25$. The last term is contribution 
of d2 Einstein gravity. As one sees it is again easy to achieve 
the coincidence of holographic CA (\ref{Ivii}) with QFT plus brane QG 
result via the identification of ${l \over \kappa^2}$ and 
${c \over 2\pi}$. It again suggests on the appearence of 
dual brane QG with matter in AdS$_3$/CFT$_2$ correspondence. 

Thus, we presented the arguments which indicate that dual brane QG
(non-conformal Einstein or conformal Weyl one) 
may also appear in AdS/CFT correspondence. Of course, more work is necessary 
to understand better the appearence of brane QG on dual QFT side 
of AdS/CFT correspondence. However, the indications we presented here look
quite promising. 

\section{AdS/CFT and quantum induced 3d brane-worlds}

It is expected that AdS/CFT correspondence \cite{AdS} should be related 
with Randall-Sundrum orbifold compactification \cite{RS} which 
presumbly is realized as warped compactification in string theory.
One scenario of this sort has been suggested in refs.\cite{NOZ,HHR0} where 
the quantum effects of brane CFT (including brane QG) have been taking into 
account. In this way 4d New Brane World may be constructed in frames of 
AdS/CFT set-up (as a kind of holographic renormalization group flow).
Moreover, such scenario is extended now for d5 gauged supergravity
 (inclusion of non-constant dilaton) where non-singular dilatonic inflationary
brane-world occurs\cite{NOO} (for related works, see\cite{ANO,HHR2,NOO1}).
The important role in the construction of such brane quantum field theory 
induced bulk Universe belongs to conformal anomaly. In the present 
section using 2d conformal anomaly we demonstrate the universality of 
scenario \cite{NOZ,HHR0} in various dimensions. In particulary,
we show that it works also for 3d inflationary (or hyperbolic) 
brane-worlds in AdS3/CFT2 correspondence.

We start with the action $S$ which is the sum of 
the three-dimensional Einstein-Hilbert action $\SEH$, 
the Gibbons-Hawking surface term $\SGH$, the surface counter 
term $S_1$ and the 2d trace anomaly induced action $W$: 
\bea
\label{Stotal2}
S&=&\SEH + \SGH + 2 S_1 + W \\
\label{SEHi}
\SEH&=&{1 \over \kappa^2}\int d^3 x \sqrt{g_{(3)}}\left(R_{(3)} 
- {2 \over l^2}\right) \\
\label{GHi}
\SGH&=&{2 \over \kappa^2}\int d^2 x \sqrt{g_{(2)}}
\nabla_\mu n^\mu \\
\label{S1}
S_1&=& {2 \over \kappa^2 }\int d^2 x \sqrt{g_{(2)}} \\
\label{W}
W&=& -{1 \over 2}\int d^2x \sqrt{-g_{(2)}} 
{N \over 48\pi}R_{(2)}{1 \over \Delta_{(2)}}R_{(2)} \ .
\eea
A solution in the bulk 3d spacetime is anti-de Sitter 
(AdS) space, whose metric is given by
\be
\label{AdS1}
ds^2= dz^2 + \e^{{2z \over l}}\sum_{i=1}^2 \left(dx^i\right)^2\ .
\ee
In (\ref{AdS1}) the slice of 
constant $z$ is a flat 2d space. One can choose, however, 
so that the slice is 2d sphere or hyperboloid. 
\bea
\label{AdS2}
ds^2 &=& dz^2 + l^2 \sinh^2 {z \over l} dS_2^2 \\
\label{AdS3}
ds^2 &=& dz^2 + l^2 \cosh^2 {z \over l} dH_2^2 \ .
\eea
Here $dS_2^2$ and $dH_2^2$ expresses the metric of unit radius 
2d sphere and hyperboloid, respectively. The metrics 
$dS_2^2$ and $dH_2^2$ can be, for example, expressed in the 
following forms:
\be
\label{AdS4}
dS_2^2={1 \over \cosh^2 \sigma}\left(d\sigma^2 + d\phi^2\right)
\ , \quad 
dH_2^2={1 \over \sinh^2 \sigma}\left(d\sigma^2 + d\phi^2\right)
\ .
\ee
Here $\phi$ has a period of $2\pi$. The metric of the flat 2d 
space is also expressed as 
\be
\label{AdS5}
\sum_{i=1}^2dx_i^2 = l^2
\e^{2\sigma}\left(d\sigma^2 + d\phi^2\right)\ .
\ee
Then all metrics in (\ref{AdS1}) 
and (\ref{AdS2}) have the following common form:
\be
\label{AdS6}
ds^2 = dz^2 + l^2\e^{A(z,\sigma)}\left(d\sigma^2 + d\phi^2\right)\ ,
\quad A(z,\sigma)=A_1(z) + A_2(\sigma)\ .
\ee
We now assume that there is a brane at $z=z_0$. The shape of the 
brane can be 2d sphere S$_2$, 2d flat space R$_2$ or 2d 
hyperboloid H$_2$ by the 
choice of the metric. 
Then by the variation of the action with respect to $A$,  one
obtains the following equations on the boundary:
\be
\label{B1}
{8 \over \kappa^2}\left( -A_{1,z}(z_0)+ {1 \over l}\right) 
+ {N \over 12\pi }k\e^{A_1(z_0)}=0\ .
\ee
Here $k=1$ for S$_2$, $k=0$ for R$_2$ and 
$k=-1$ for H$_2$. Note that 
$A_1(z)=\ln\cosh{z \over l}$ for S$_2$, 
$A_1(z)={z \over l}$ for R$_2$ and 
$A_1(z)=\ln\sinh{z \over l}$ for H$_2$. 
For 2d flat space, Eq.(\ref{B1}) becomes identity. For 
2d sphere, Eq.(\ref{B1}) has the following form:
\be
\label{B2}
{8 \over \kappa^2 l}\left(\coth{z_0 \over l} - 1 \right)
\sinh^2{z_0 \over l}={N \over 12\pi}
\ee
and for 2d hyperboloid, Eq.(\ref{B1}) has the following form:
\be
\label{B3}
{8 \over \kappa^2 l}\left(1 - \tanh{z_0 \over l} \right)
\cosh^2{z_0 \over l}={N \over 12\pi}\ .
\ee
Both of the equations (\ref{B2}) and (\ref{B3}) can be 
uniquely solved with respect to $z_0$. This situation is 
different from the situation in 4d quantum induced curved brane\cite{NOZ}
 in AdS$_5$,
 where 
there is no usually solution for the brane of the shape of 
the hyperboloid. The difference appears since the trace anomaly 
is linear on the scalar curvature in 2 dimensions 
but the anomaly is proportional to the squares of the curvatures 
in 4 dimensions. In 2 dimensions, the change of the sign coming 
from $k$ can be absorbed into the change of sign in the 
scalar curvature (the scalar curvature is positive for 2d 
sphere but negative for hyperboloid). 
Since the radius $R$ of the brane can be defined by 
\be
\label{B4}
R=l\e^{A_1(z_0)}\ ,
\ee
Eqs.(\ref{B2}) and (\ref{B3}) can be rewritten in the 
following forms, respectively:
\bea
\label{B5}
{8R \over \kappa^2 l^2}\left(\sqrt{ {R^2 \over l^2} 
+ 1 } - {R \over l} \right) &=& {N \over 12} \\
\label{B6}
{8R \over \kappa^2 l^2}\left({R \over l} - \sqrt{ 
{R^2 \over l^2} - 1 }  \right) &=& {N \over 12} \ .
\eea
Both of the l.h.s.'s in Eqs.(\ref{B5}) and (\ref{B6}) are 
monotonically increasing functions. Since the l.h.s. in 
(\ref{B5}) vanishes when ${\cal R}=0$ and goes to positive 
infinity when ${\cal R}\rightarrow \infty$, (\ref{B5}) 
determines ${\cal R}$ uniquely. 
On the other hand, the l.h.s. in (\ref{B6}) 
becomes ${8 \over \kappa^2 l}$ when $R=l$. Therefore if 
\be
\label{B7}
{8 \over \kappa^2 l}<{N \over 12} \ ,
\ee
there is a non-trivial solution of the hyperbolic brane. 
Thus, we demonstrated the possibility of inflationary or hyperbolic 
3d brane-world induced by quantum effects of brane matter. As we saw in previous
section the 2d conformal anomaly may include also quantum contribution from
2d gravity (then sign of quantum effective action $W$ may be positive,
depending on the amount of matter). This indicates that scenario of
refs.\cite{NOZ,HHR0} is quite universal. Moreover, in the same way as 
in ref.\cite{NOO} it may be easily extended for the presence of non-trivial
dilaton(s). 

\section{Quantum-corrected energy and quantum entropy bounds in FRW-Universe}

Holographic principle suggests the interesting bounds between 
microscopic entropy and Bekenstein-Hawking 
entropy \cite{hawking} as it was suggested in refs.\cite{hooft}.
This indicates also to the relations between AdS/CFT set-up 
and entropy and brings new bounds for the entropy \cite{bousso,maldacena,
bekenstein} (for related works on holographic entropy, see\cite{entropy}).
Note that 
it would be really interesting to understand the entropy bounds origin
from stringy (AdS/CFT) points of view (for recent attempt to understand 
cosmology/CFT set-up via corresponding comparison of entropies, see 
\cite{cosm}).

In the present section we make the attempt to understand the role
of (large) quantum corrections to energy/entropy (using again the 
conformal anomaly ) in the study of entropy bounds and 
their dynamical evolution \cite{EV}( for related works, see\cite{related}).
 The dynamical equations describing
FRW-cosmology and energy/entropy are defined in the presence 
 of quantum corrections which could be dominant ones. Our analysis 
suggests that the effect of (large) quantum contribution may destroy
the well-accepted entropy bounds which become the approximate ones 
and which occur only for some time-limited period of Universe evolution.

We start with the following action:
\bea
\label{Stotal}
S&=&{1 \over 16\pi G}\int d^4 x \sqrt{-g}R_{(4)} 
+ S_{\rm matter} + W \nn
W&=& b \int d^4x \sqrt{-\widetilde g}\widetilde F A \nn
&& + b' \int d^4x \sqrt{-\widetilde g}
\left\{A \left[2{\widetilde\Box}^2 
+\widetilde R_{\mu\nu}\widetilde\nabla_\mu\widetilde\nabla_\nu 
 - {4 \over 3}\widetilde R \widetilde\Box^2 
+ {2 \over 3}(\widetilde\nabla^\mu \widetilde R)\widetilde\nabla_\mu
\right]A \right. \nn
&& \left. + \left(\widetilde G - {2 \over 3}\widetilde\Box \widetilde R
\right)A \right\} \nn
&& -{1 \over 12}\left\{b''+ {2 \over 3}(b + b')\right\}
\int d^4x \sqrt{-\widetilde g}
 \left[ \widetilde R - 6\widetilde\Box A 
 - 6 (\widetilde\nabla_\mu A)(\widetilde \nabla^\mu A)
\right]^2 \nn
&& +S_{\rm inv} \ .
\eea 
The last term ($W$) represents the conformal anomaly induced effective 
action (for a review, see \cite{BOS}).
Here one  chooses the metric in the following form:

\be
\label{en1}
ds^2 = \e^{2A(\tau)}\tilde g_{\mu\nu}dx^\mu dx^\nu
\equiv \e^{2A(\tau)}\left(-d\tau^2 + d\Omega_3^2\right)
\ee
 $d\Omega_3^2$ expresses the metric of 3 dimensional sphere.
We denote the $A$ independent terms by $S_{\rm inv}$ which would 
be given as 
\be
\label{en1b}
S_{\rm Inv}= \int d^4x \sqrt{-\widetilde g}\left(
a_1 \widetilde R^2 
+ a_2 \widetilde R_{\mu\nu}\widetilde R^{\mu\nu}
+ a_3 \widetilde R_{\mu\nu\rho\sigma}
\widetilde R^{\mu\nu\rho\sigma}\right) 
+ \mbox{non-local terms}\ .
\ee
It is clear that such terms (Casimir energy of static space) 
cannot be obtained by only the integration of conformal anomaly. 

 Let us neglect the non-local terms in (\ref{en1b}) and 
choose the metric in (\ref{en1}) as (\ref{en1b}) then

\be
\label{en1bb}
S_{\rm Inv}= \tilde a \int d^4x \sqrt{\widetilde g}\ ,
\quad \tilde a = 36 a_1 + 12 a_2 + 12 a_3\ .
\ee

By the variation over $\tilde g_{\tau\tau}$, we obtain the 
following equation (conservation law):
\bea
\label{en2}
0&=&{1 \over 16\pi G}\e^{2A}\left\{6 
+ 6 \left(\partial_\tau A\right)^2\right\} 
+ 2\e^{6A}T^{\tau\tau}_{\rm matter} \nn
&& + b'\left\{4\partial_\tau A\partial_\tau^3 A 
 -2\left(\partial_\tau^2 A\right)^2 
 +8\left(\partial_\tau A\right)^2 \right\} \nn
&& - {1 \over 12}\left\{b'' + {2 \over 3}\left(b+b'\right)
\right\}\left\{ - 36\left(\partial_\tau^2 A\right)^2
 -108 \left(\partial_\tau A\right)^4 \right. \nn
&& \left.+ 72 \partial_\tau A \partial_\tau^3 A 
 - 72 \left(\partial_\tau A\right)^2 + 36\right\} + \tilde a \ .
\eea
Here $T^{\tau\tau}$ is the stress energy tensor of the matter 
\be
\label{en3}
T^{\tau\tau}=-{2 \over \sqrt{-g}}{\delta S_{\rm matter} \over 
\delta g_{\tau\tau}}\ .
\ee
On the other hand, varying over $A$ one gets the dynamical field equation
\bea
\label{en4}
0&=&{1 \over 16\pi G}\e^{2A}\left\{12 
+ 12 \partial_\tau^2 A
+ 12 \left(\partial_\tau A\right)^2\right\} \nn
&& + b'\left\{4\partial_\tau^4 A +16 \partial_\tau^2 A \right\} \nn
&& - \left\{b'' + {2 \over 3}\left(b+b'\right)
\right\}\left\{ 6\partial_\tau^6 A 
 -36 \left(\partial_\tau A\right)^2 \partial_\tau^2 A 
 -12 \partial_\tau^2 A \right\}\ .
\eea
 Our assumption is the matter is conformally invariant , that is, 
its classical stress-energy tensor is traceless. Then the matter 
 does not contribute to the above equation (\ref{en4}). 
We now change the time coordinate from the conformal one $\tau$ 
to the cosmological one $t$ by
\be
\label{en5}
dt = \e^A d\tau\ .
\ee
 One also defines the radius of the universe $R$ and 
the Hubble constant $H$ as follows:
\be
\label{en6}
R\equiv\e^A\ ,\quad H={1 \over R}{d R \over dt}
={d A \over dt}\ .
\ee
Then Eqs.(\ref{en2}) and (\ref{en4}) are rewritten as follows:
\bea
\label{en7}
H^2&=& - {1 \over R^2} + {16\pi G \over 6}{E \over V} \\
\label{en8}
{E \over V}&\equiv& T^{tt} 
 - b'\left(4H H_{,tt} + 12 H_{,t} H^2 - 2H_{,t}^2 + 6H^4 
+ {8 \over R^2} H^2\right) \nn
&& +{1 \over 12}\left\{b'' + {2 \over 3}\left(b+b'\right)
\right\} \nn
&& \quad \times \left(- 36 H_{,t}^2 + 216 H_{,t} H^2 
+ 72H H_{,tt} - {72 \over R^2}H^2+{36 \over R^4} \right) \nn
&& + {\tilde a \over R^4} \ ,\\
\label{en7bb}
0&=&{ 1 \over 16\pi G}\left({12 \over R^2} + 24 H^2 + 12 H_{,t}
\right) \nn
&& + b'\Bigl\{4H_{,ttt} + 28 H H_{,tt} + 16 H_{,t}^2 
+ 72 H^2 H_{,t} \nn 
&& + 24 H^4 + {16 \over R^2}\left(H_{,t} + H^2 \right)\Bigr\} \nn
&& -\left\{b'' + {2 \over 3}\left(b+b'\right)
\right\}\Bigl\{6 H_{,ttt} + 42 H H_{,tt} + 24 H_{,t}^2 
+ 72 H^2 H_{,t}\nn
&& - {12 \over R^2}\left(H_{,t} + H^2\right)
\Bigr\}\ .
\eea
Here
\be
\label{en7b}
T^{tt}=\e^{2A}T^{\tau\tau}\ .
\ee
Especially when $b=b'=b''=0$, the equation (\ref{en7bb}) has the 
following form:
\be
\label{en7bbb}
0={12 \over R^2} + 24 H^2 + 12 H_{,t}\ .
\ee
If  one deletes $H^2$ term from (\ref{en7bbb}) 
by using (\ref{en7}), then
\be
\label{en7bbb2}
H_{,t}={1 \over R^2} - {16\pi G \over 3}{E \over V}\ ,
\ee
 This is the standard evolution equation of Einstein theory 
for the conformal matter, where $3p=E$ ($p$ is the pressure).
The important remark is in order. In the above formalism 
the matter is considered to be classical and only classical stress-tensor 
appears. The quantum effects of matter (and of quantum gravity) via
the corresponding anomaly induced effective action $W$ modify the 
classical general relativity by extra (non-local) terms. Hence,
quantum effects are accounted here via the modification of 
classical gravity.

If one defines the Bekenstein-Hawking entropy ${\cal S}_{\rm BH}$, 
the Bekenstein entropy ${\cal S}_{\rm B}$ and the Hubble 
entropy ${\cal S}_{\rm H}$ as
\be
\label{en9}
{\cal S}_{\rm BH}={1 \over 2}{V \over GR}\ ,\quad 
{\cal S}_{\rm B}={2 \over 3}\pi ER\ ,\quad 
{\cal S}_{\rm H}={2HV \over 4G}\ ,
\ee
Eq.(\ref{en7}) leads to
\be
\label{en10}
{\cal S}_{\rm H}^2 + \left({\cal S}_{\rm BH}
 - {\cal S}_{\rm B}\right)^2 
={\cal S}_{\rm B}^2\ .
\ee

When $T^{tt}=0$,  the solution in the form of de Sitter space exists 
\cite{S,MM}
\be
\label{en11}
R=A\cosh B t \ ,\quad A,B=\mbox{constant}\ ,
\ee
 Using (\ref{en11}) in (\ref{en7}) and (\ref{en8}), 
we find that 
(\ref{en11}) is solution if 
\be
\label{en12}
B^2={1 \over  A^2}=- {1 \over 16\pi G b'}\ .
\ee
This is the basis for anomaly driven inflation \cite{S} (
for recent extended arguments in favor of realistic anomaly-driven inflation
 ,see \cite{HHR2}).
When one substitutes Eqs.(\ref{en11}) and (\ref{en12}), Eq.
(\ref{en7}) is satisfied if
\be
\label{en13}
\tilde a = -8b'\ .
\ee
In the following, we assume $\tilde a$ is given by (\ref{en13}) (the choice 
of normalization). 
The obtained metric of de Sitter space has the following 
form:
\be
\label{dS1}
ds^2= - dt^2 + A^2 \cosh^2 {t \over A}d\Omega_3^2\ .
\ee
If we analytically continue $t$ as
\be
\label{dS2}
t=iA\left(\theta - {\pi \over 2}\right)\ ,
\ee
we obtain the metric of 4 dimensional sphere S$_4$:
\be
\label{dS3}
ds^2 = A^2\left(d\theta^2 + \sin^2\theta d\Omega_3^2\right)\ .
\ee
 Identifying the north pole ($\theta=0$) and the south pole 
($\theta=\pi$), the period of $\theta$ is $\pi$. Then one could 
regard that $t$ has a period ${\pi \over A}$, which is related with  
the inverse temperature ${\cal T}$
\be
\label{dS4}
{\cal T}={A \over \pi}\ .
\ee

In the above discussion we used the approach where quantum effects modify 
the gravitational effective action. One can show that it is equivalent 
to the approach where quantum corrections to stress-energy tensor are 
considered.
Indeed in \cite{HHR2} there has been given an elegant method to 
obtain the (quantum corrected) energy in the spacetime with the metric:
\be
\label{hhr1}
ds^2 = d\sigma^2 + b(\sigma)^2 d\Omega_3^2\ .
\ee
The authors of ref.\cite{HHR2} worked in the Euclidean metric in 
order to consider instanton solutions. Then one can identify 
\be
\label{hhr2}
\sigma = it\ ,\quad b(\sigma)=R(t)\ .
\ee
Here $t$ and $R(t)$ were given in (\ref{en5}) and (\ref{en8}), 
respectively. In the following, we write $b(\sigma)$ as $R$ in 
order to avoid the confusion with $b$, $b'$ and $b''$ in 
(\ref{Stotal}) and only consider the Lorentzian signature 
case.  In \cite{HHR2}, there was found the quantum energy for 
${\cal N}=4$ $SU(N)$ or $U(N)$ supersymmetric Yang-Mills (SYM)
theory, where
\be
\label{hhr7}
b=-b'= {N^2 \over 4(4\pi)^2}\ ,\quad b''=0\ ,
\ee 
Here we generalize the method to more general $b$, $b'$ and 
$b''$ case. It is interesting that formally one can include 
the quantum gravity corrections to coefficients of conformal
anomaly (see Section 2), then quantum stress-energy tensor 
is given also by sum of two parts: matter and gravity.
We now define the energy density $\rho$ and pressure $p$ by 
\be
\label{hhr4}
\left\langle T_{tt}\right\rangle =\rho\ ,\quad
\left\langle T_{ij}\right\rangle =p g_{ij} 
= p\e^{2A}\tilde g_{ij}\ (i,j=1,2,3)\ ,
\ee
 Then
\be
\label{hhr5}
\left\langle T\right\rangle \equiv g^{\mu\nu}
\left\langle T_{\mu\nu}\right\rangle=-\rho + 3p \ .
\ee
On the other hand, the conservation law of the 
energy-momentum tensor $\nabla^\mu T_{\mu\nu}=0$ tells 
\be
\label{hhr6}
0=\rho_{,t} + 3A_{,t} (\rho + p)\ .
\ee
Combining (\ref{hhr5}) and (\ref{hhr6}), one gets
\be
\label{hhrA1}
\left(\e^{4A}\rho \right)_{,t}=-\e^{4A}A_{,t}
\left\langle T\right\rangle\ .
\ee
Since now the trace anomaly is given by
\bea
\label{hhrA2}
\left\langle T\right\rangle&=& b\left( F + {2 \over 3}
\Box {\cal R}\right) + b' G + b'' \Box {\cal R} \nn
&=& b'\e^{-3A}\left\{ 8\left(\e^{3A}A_{,t}^3 \right)_{,t} 
+ 24 \left(\e^A A_{,t}\right)_{,t} \right\} \nn 
&& - \left({2 \over 3}b + b''\right)\e^{-3A}\left\{ 
\e^{3A}\left(6 A_{,tt} + 12 A_{,t}^2 + 6\e^{-2A}\right)_{,t}
\right\}_{,t}\ ,
\eea
 one finds
\bea
\label{hhrA3}
\rho&=&-\left.{1 \over R^4}\right[b'\left( 6 R^4 H^4 
+ 12 R^2 H^2\right) \nn
&& \left. + \left({2 \over 3}b + b''\right)\left\{ R^4
\left( -6 H H_{,tt}- 18 H^2 H_{,t} + 3 H_{,t}^2 \right) 
+ 6R^2 H^2\right\}  + C \right]\ .
\eea
Here $C$ is the integration constant. If we choose 
the constant $C$ as 
\bea
\label{hhrA4}
C=-3 b'' - 2(b+b') - \tilde a = -2b +6 b' -3b''\ , 
\eea
we find that $\rho$ coincides with ${E \over V}$ in (\ref{en8}) 
when there is no contribution from the classical matter 
($T_{tt}=0$). This proves the equivalency of two formalisms to
derive the evolution equations.
 
 One can also obtain the expression of the pressure $p$ from 
(\ref{hhr5}), (\ref{hhrA2}) and (\ref{hhrA3}):
\bea
\label{hhrAA1}
p&=&b'\left\{ 6 H^4 + 8H^2 H_{,t} + {1 \over R^2}\left( 
4H^2 + 8 H_{,t}\right) \right\} \nn
&& \left.+ \left({2 \over 3}b + b''\right)\right\{ -2H_{,ttt} -12 
H H_{,tt} - 18 H^2 H_{,t} - 9 H_{,t}^2 \nn
&& \left. + {1 \over R^2}
\left( 2H^2 + 4H_{,t}\right) \right\}  - {C \over 3R^4}\ .
\eea

 One can now divide ${E \over V}$ in (\ref{en8}) or $\rho$ in 
(\ref{hhrA3}) and $p$ in (\ref{hhrAA1}) with the contribution 
from the classical matter 
by the $R$ dependence as follows
\bea
\label{Cs1}
\rho&=&\rho_0 + \rho_1 + \rho_2 \nn
\rho_0&=&T_{tt} -6b' H^4 - \left({2 \over 3}b + b''\right)
\left( -6 H H_{,tt}- 18 H^2 H_{,t} + 3 H_{,t}^2 \right) \nn
\rho_1&=&- \left\{12b' + 6 \left({2 \over 3}b + b''\right)\right\}
 {H^2 \over R^2} \nn
\rho_2&=& - {C \over R^4}\\
\label{Cs1b}
p&=&p_0 + p_1 + p_2 \nn
p_0&=&{1 \over 3}T_{tt} + b'\left( 6 H^4 + 8H^2 H_{,t} \right) \nn
&& + \left({2 \over 3}b + b''\right)\left( -2H_{,ttt} -12 
H H_{,tt} - 18 H^2 H_{,t} - 9 H_{,t}^2\right) \nn
p_1&=&{1 \over R^2}\left\{b'\left( 
4H^2 + 8 H_{,t}\right)  + \left({2 \over 3}b + b''\right) 
\left( 2H^2 + 4H_{,t}\right) \right\}  \nn
p_2&=& - {C \over 3R^4}\ .
\eea
$\rho_0$ does not depend on $R$, therefore $\rho_0$ expresses the 
extensive part of the energy. On the other hand, $\rho_1$ and 
$\rho_2$ will express the finite size effects (like the Casimir 
energy). 
Using the thermodynamical equation for the entropy 
${\cal S}$ 
\be
\label{ETS}
E=-pV + {\cal T}{\cal S}\ ,
\ee
we naively find the entropy ${\cal S}$ is given by
\be
\label{ETS2}
{\cal S}={V(\rho + p) \over {\cal T}}\ .
\ee
It is very important to realize that quantum contributions in
above expression (as well as for energy) may be dominant if compare 
with classical expression!

Let us consider the case of the de Sitter space solution in 
(\ref{dS1}) with (\ref{en12}). In this case, the quantum energy density 
$\rho$ (\ref{hhrA3}) and the quantum pressure $p$ (\ref{hhrAA1}) 
are given by
\be
\label{prho1}
\rho=-p=- {6b' \over A^4}
-6b' B^4=- {6 \over \left(16\pi G\right)^2 b'}\ .
\ee
This indicates that quantum entropy ${\cal S}$ vanishes (from 
(\ref{ETS2})) if the temperature is finite as in (\ref{dS4}).
This might tell that the definition of the entropy (if the 
temperature ${\cal T}$ is defined) is not correct. 
Usually the entropy is extensive quantity, i.e., proportional 
to the volume of the space. Eqs.(\ref{Cs1}) and (\ref{Cs1b}) 
might, however, tell that the definition in (\ref{ETS2})  is
 a mixture of the extensive part and non-extensive part. 

In \cite{EV}, another better way to define the entropy, which is 
extensive, has been proposed. 
The energy $E=\rho V$ is divided to the extensive part $E_E$ and 
the Casimir (or quantum) part $E_C$:
\be
\label{EV0}
E=E_E + {1 \over 2}E_C\ . 
\ee
If  one introduces a length parameter $l$, 
$E_E$ scales as $E_E\sim l^3 $ and $E_C$ as $E_C\sim l$. 
If there is a conformal symmetry, 
both of $RE_E$ and $RE_C$ should only depend on the entropy 
${\cal S}$. If the entropy ${\cal S}$ is extensive quantity 
and scales as ${\cal S}\sim l^3$,  one finds
\be
\label{EV1}
RE_E\propto {\cal S}^{4 \over 3}\ , \quad 
RE_C\propto {\cal S}^{2 \over 3}\ .
\ee
Therefore
\be
\label{EV2}
{\cal S} \propto \sqrt{E_E E_C}\ .
\ee
In \cite{EV}, the constant of the proportionality has been 
determined from the AdS/CFT correspondence as
\be
\label{EV3}
{\cal S}={2\pi R \over 3}\sqrt{2E_E E_C}\ .
\ee
By using (\ref{Cs1}), we might be able to identify
\be
\label{EV4}
E_E = \rho_0 V\ ,\quad E_C=2\rho_1 V
\ee
and we have 
\be
\label{EV5}
{\cal S}={4\pi RV \over 3}\sqrt{\rho_0 \rho_1}\ .
\ee
Here we dropped the integration constant dependent $\rho_2$  
because the entropy would not become 
extensive if we include $\rho_2$ since $\rho_2$ is expected 
to scale like $l^{-1}$. 
Since $b''$ can be shifted by the finite regularization, we 
choose $b''=-{2 \over 3}b$  for simplicity. Then from 
(\ref{Cs1}), one gets
\be
\label{EV6}
{\cal S}=-8\pi \sqrt{2}V b' H^3\ .
\ee
Here we put the energy density of the classical 
matter to vanish $T_{tt}=0$. 
We now consider the solution of the de Sitter space in 
(\ref{dS1}) with (\ref{en12}) and compare the entropy 
in (\ref{EV6}) with the entropies defined in (\ref{en9}). 
 Using de Sitter solution in  the expression 
 for the entropies, one gets
\bea
\label{EV7}
{\cal S}&=& - 8\pi \sqrt{2} Vb' B^3 \tanh^3 Bt \nn
{\cal S}_{BH}&=& -{8\pi Vb' B^3 \over \cosh Bt} \nn
{\cal S}_B&=& -4 \pi Vb' B^3 \cosh Bt \nn
{\cal S}_H&=& - 8\pi Vb' B^3 \tanh Bt \ .
\eea
In (\ref{EV7}), ${\cal S}$, ${\cal S_B}$ and ${\cal S}_H$ are 
monotonicaly increasing functions of $t$ when $t\geq 0$. 
${\cal S}$ and ${\cal S}_H$ vanish and ${\cal S}_B$ has a 
minimum value $-4 \pi Vb' B^3$ when $t=0$ and ${\cal S}$ 
and ${\cal S}_H$ take a 
maximum values, $- 8\pi \sqrt{2} Vb' B^3$ and 
$- 8\pi Vb' B^3$, respectively, and ${\cal S}_B$ 
increases exponentially when $t\rightarrow +\infty$. 
On the other hand, ${\cal S}_{BH}$ is a monotonically decreasing function 
of $t$ when $t\geq 0$. ${\cal S}_{BH}$ takes a maximum value 
$-8\pi Vb' B^3$ when $t=0$ and vanishes when 
$t\rightarrow + \infty$. 
${\cal S}_B$, ${\cal S_H}$ and ${\cal S}_{BH}$ coincide with 
each other when $\tanh Bt=\tanh Bt_1
={1 \over \sqrt{2}}=0.7071...$.  One finds ${\cal S_{BH}}>
{\cal S}_B>{\cal S}_H$ when $t<t_1$ and ${\cal S}_B>
{\cal S}_H>{\cal S_{BH}}$ when $t>t_1$. We should note that 
${\cal S}_B$ does not cross with ${\cal S}_H$ but 
${\cal S}_B$ is tangent to ${\cal S}_H$ when $t=t_1$. 
${\cal S}$ crosses with ${\cal S}_{BH}$ when $\tanh Bt=\tanh Bt_2
= 2^{-{1 \over 4}}=0.7679...$ and with ${\cal S}_H$ 
when $\tanh Bt=\tanh Bt_3= 0.84089...$. By summarizing the above 
behavior of the entropies, one gets
\bea
\label{smry}
&{\cal S_{BH}}>{\cal S}_B>{\cal S}_H>{\cal S}\quad &t<t_1 \nn
&{\cal S}_B>{\cal S}_H>{\cal S_{BH}}>{\cal S}\quad &t_1<t<t_2 \nn
&{\cal S}_B>{\cal S}_H>{\cal S}>{\cal S_{BH}}\quad &t_2<t<t_3 \nn
&{\cal S}_B>{\cal S}>{\cal S}_H>{\cal S_{BH}}\quad &t>t_3 \ .
\eea
The entropy bounds ${\cal S}<{\cal S}_{BH}$ 
and ${\cal S}<{\cal S}_H$ are valid only for 
small $t$. On the other hand, the bound ${\cal S}<{\cal S}_B$ 
is valid only 
for large $t$. (Here we assume $b'$ is negative as usually occurs even
in the presence of quantum gravity.) 
Therefore the entropy bounds seem to break down in general due 
to the quantum effects. This indicates that in the region where 
quantum effects are dominant, one should re-consider 
the fundamental physical laws as it was suggested some time ago 
by 't Hooft.

\section{Discussion}.

In summary, we were trying to clarify the role and the importance of 
conformal anomaly in AdS/CFT correspondence and in related issues.
In particulary, the indication to the window for realization of brane 
quantum gravity in AdS/CFT set-up (via comparison of holographic and QFT 
conformal anomalies) is presented. The occurence of quantum induced 
inflationary (or hyperbolic) brane-world scenario in terms of AdS3/CFT2 
correspondence is presented. Quantum corrected entropy for inflationary 
Universe is proposed and some related entropy bounds which turned out to
be evolving are discussed. Again, the basic elements of this calculation
 are the conformal anomaly and AdS/CFT.

Of course, our study being useful in clarification of some aspects of
AdS/CFT may be extended in various directions.
For example, the better understanding of 
incorporation of Randall-Sundrum scenario to string theory is required.
New Brane World realized in third section in frames of holographic flow
 within AdS3/CFT2 correspondence provides the useful background for such 
 understanding. From another side, the hypotetical
 quantum corrected entropy bounds
  are estimated only for inflationary Universe (its early and late stages).
It is clear that at the exit from inflationary stage these bounds should be 
completely modified. It would be really interesting to study their evolution
 especially at the end of the inflation.
Moreover, it could be that even the definition of quantum entropy 
at early Universe should be modified. This fundamental problem deserves 
 the very careful future investigation.
\ 

\noindent
{\bf Acknoweledgements}

We are very grateful to G. Gibbons, O. Obregon and
 V.I. Tkach for useful discussions.
The work by SDO has been supported in part by CONACyT (CP, Ref.990356 and
grant 28454E) and in part by RFBR grant N99-02-16617.

\end{document}